
\input harvmac

\def\Ms{M(\sigma)}

\def\es{\epsilon(\sigma)}
\def\pder#1{{\partial\over\partial #1}}
\def\dpder#1{{\partial^2\over\partial{#1}^2}}
\def\proj#1{{\cal P}_{#1}}
\def\bra{\langle0|}
\def\ket{|0\rangle}
\def\lslash#1{{#1}\llap/}
\def\dslash#1{\rlap{$\,$/}{{#1}\llap/}}
\def\uslash#1{\underline{#1}\llap/}
\def\absig{(|\sigma|+|\sigma'|)}

\def\ptensor#1{T_{#1}^{\mu\nu\rho\sigma}}

\def\sitarel#1#2{\mathrel{\mathop{\kern0pt #1}\limits_{#2}}}
\def\IE{{\it i.e.}}
\def\T{{\rm T}}
\def\DENNO#1{{1 \over [M^2+\alpha\omega^2+(1-\alpha){\omega'}^2
                                          -\alpha(1-\alpha)p^2]^{#1}}}
\def\DENO#1{[M^2+\alpha\omega^2+(1-\alpha){\omega'}^2
                                          -\alpha(1-\alpha)p^2]^{#1}}
\def\DENOO{[M^2+\alpha\omega^2+(1-\alpha){\omega'}^2
                                          -\alpha(1-\alpha)p^2]}

\def\deno#1{[M^2+\alpha\omega^2+(1-\alpha){\omega'}^2]^{#1}}
\def\DT{{D \over 2}}
\def\NPBsupl#1#2#3{Nucl. Phys. {\bf B} (Proc. Suppl.) {\bf #1} (#2) #3}
\def\NPB#1#2#3{Nucl. Phys. {\bf B #1} (#2) #3}
\def\PLB#1#2#3{Phys. Lett.  {\bf B #1} (#2) #3}

\def\PRL#1#2#3{Phys. Rev. Lett. {\bf  #1} (#2) #3}

\def\PRL#1#2#3{Phys. Rev. Lett. {\bf  #1} (#2) #3}

\font\fourteenrm=cmr12 scaled \magstep1
\font\twelverm=cmr12

\rightline{\vbox{
                  \hbox{KUNS-1239}
                  \hbox{HE(TH)94/01}
                 \hbox{February, 1994} } }
\vskip 24pt

\vskip 24pt
\centerline{
\fourteenrm
\baselineskip=20pt
\vbox{
\hbox{
On the Large Mass Limit of the Continuum Theories
}
\centerline{\hbox{
in Kaplan's Formulation
}}
}}
\vskip 36pt
\vskip 36pt
\centerline{
\twelverm
Teruhiko Kawano
           \footnote{$^{\dag}$}{e-mail: kawano@gauge.scphys.kyoto-u.ac.jp}
 and
Yoshio Kikukawa
           \footnote{$^{\ddag}$}{e-mail: kikukawa@gauge.scphys.kyoto-u.ac.jp} }
\vskip 12pt
\centerline{\twelverm Department of Physics, Kyoto University}
\centerline{\twelverm Kyoto 606-01, Japan}

\vskip 32pt

\noindent
\centerline{\bf Abstract}
   Being inspired by Kaplan's proposal for simulating chiral fermions
  on a lattice, we examine the continuum analog of his domain-wall
  construction for two-dimensional chiral Schwinger models.
  Adopting slightly unusual dimensional regularization, we explicitly
  evaluate the one-loop effective action in the limit that the
  domain-wall mass goes to infinity. For anomaly-free cases, the
  effective action turns out to be gauge invariant in two-dimensional sense.
\par
{\leftskip 0.4 in \rightskip 0.4 in
\par}
\vfill
\eject


\newsec{Introduction}

   The ($2n+1$)-dimensional Dirac fermion with the domain-wall mass
  has interesting properties as discussed by Callan and Harvey
\ref\CalHav{C.G. Callan Jr. and J.A. Harvey, \NPB {250} {1985} {427}.}.
  Here the domain-wall mass means the mass depending on
  the extra coordinate $\sigma$: it behaves asymptotically like
  $M(\sigma) \rightarrow \pm M$ as $\sigma \rightarrow \pm \infty$
  and vanishes at $\sigma =0 $. A $2n$-dimensional chiral fermion
  can arise from the Dirac fermion and is bound to the domain wall
  \IE\ the $2n$-dimensional hyperplane specified by $\sigma = 0$.
  It was argued that, when the $(2n+1)$-dimensional fermion is coupled
  to external gauge fields, the gauge anomaly produced by this
  chiral mode is compensated with the current flow of the massive
  modes and the gauge invariance is kept intact as a whole.

   The use of such chiral mode on the domain wall was recently proposed
  by Kaplan
\ref\kaplan{D.B. Kaplan, \PLB{288}{1992}{342}.}
  to simulate chiral fermions on a lattice, which has been
  a long-standing problem because of the species doubling
\ref\ksmit{L. Karsten and J. Smit, \NPB{183}{1981}{103}.}
\ref\nn{ H.B. Nielsen and M. Ninomiya, \NPB{185}{1981}{20};
                                                         \NPB{193}{1981}{173}}
\ref\karsten{L. Karsten, \PLB{104}{1981}{315}.}.
  In fact,  putting na{\"\i}vely the Dirac fermion on a lattice,
  we find not only the chiral mode but also their doublers on
  the domain wall.
  In this case, however, we can introduce a Wilson term for the
  ($2n+1$)-dimensional Dirac fermion in a gauge invariant way.
  Then it is possible to show that such doublers can be removed
  at least for free theories. More recently, the lattice analog
  of the Callan-Harvey analysis
\ref\jansen{K. Jansen, \PLB{288}{1992}{348}; {\bf B}296 (1992) 347.}
\ref\GJK{M.F.L. Golterman, K. Jansen and D.B. Kaplan, \PLB{301}{1993}{219}.}
  shows that the same mechanism of the anomaly cancellation also works
  in the lattice setup even though the detail of the cancellation depends on
  the ratio between the Wilson coupling and the domain wall height $M$.

  Although it is a clever idea, there are several questions for the proposal.
  First of all, it is not clear how to reduce the freedom of the
  $(2n+1)$-dimensional gauge bosons to $2n$-dimensional ones.
  The second one is how to separate the chiral zero-mode of the
  opposite chirality on the anti domain wall. Such an extra mode
  exists because of the periodicity condition on the compactified
  extra dimension which is a necessary condition practically to
  perform numerical simulation. The third question is, besides the
  anomaly cancellation, how the massive modes including doublers
  affect the low energy theory, \IE\ what about the decoupling of such modes.

   The first and second questions are rather serious.
  Since the gauge field can propagate in the extra direction
  and feel both of chiralities, we are afraid that this theory may
  become vector-like. Especially, this is actually the case, in
  particular, if we make the gauge field independent of the extra
  coordinate $\sigma$. Thus, keeping the dependence of the gauge
  field on $\sigma$, we should carefully introduce the gauge
  field\foot{An alternative approach have been proposed by Narayanan
  and Neuberger\ref\NARA{R.~Narayanan and H.~Neuberger, \PLB{302}{1993}{62};
  \NPB{412}{1994}{574}; \PRL{71}{1993}{3251}; RU-93-52,
  IASSNS-HEP-93/74.}, where the gauge field is $2n$-dimensional one
  and the extra dimension are not compactified. They have shown that
  their overlap formula reproduces the correct chiral anomaly and
  the effect of instanton.}\llap.
  In order to do so, we can conceive of at least two
  possibilities: one is that we prevent the gauge field from
  propagating in the extra direction.
  The other is that we make the gauge field couple only
  to the chiral zero-mode on the wall but not to the one on the anti-wall.
  This also needs an additional scalar field for the gauge invariance.

   The first possibility can be realized,
  if the gauge coupling $\beta_\sigma$ in the extra direction goes to
  zero. Then all the four-dimensional gauge bosons at every $\sigma$
  become independent each other. The gauge field coupled to the chiral
  zero-mode on the wall is different from that coupled to the chiral
  zero-mode on the anti-wall. But for small $\beta_\sigma$,
  in the mean-field approximation
\ref\altes{ C.P. Korthals-Altes, S. Nicolis and J. Prades,
                                                     \PLB {316}{1993}{339}.}
  there emerges the layered phase and the fermion is entirely confined
  to the four-dimensional layers. Then the fermion propagator is found
  to be vector-like.

   The latter possibility was discussed in
\ref\fifthgauge{D.B. Kaplan, \NPBsupl {30}{1993}{597}.},
  and was investigated in detail by the authors of
\ref\gjpv{M. Golterman, K. Jansen, D.N. Petcher and J.C. Vink,
                                UCSD-PTH-93-28, Sep 1993. hep-lat/9309015.}.
  They concluded that there does not exist the phase where the mirror
  mode on the waveguide could decouple. This result is also disappointing.

   As for the third question, an example of the effect of
  doublers can be seen in \jansen\GJK \ that
  the number of chiral fermions on the domain wall depends on
  the ratio between the Wilson coupling and the domain-wall height $M$.
  Aoki and Hirose
\ref\AokHir{S. Aoki and H. Hirose, UTEP-262, University of Tsukuba
                                  preprint, September 1993. hep-lat/9309014.}
  have evaluated the one-loop effective action for
  the domain-wall fermions in the would-be two-dimensional chiral
  Schwinger models and found a mass-like term for the gauge boson.
  As far as the gauge field depends on the extra coordinate $\sigma$,
  such a term exists. This term might indicate that, in the resultant
  low energy theory which we expect to be our target theory, we cannot
  obtain the two-dimensional gauge invariance even in an anomaly-free
  chiral gauge theory. Certainly such invariance in the two-dimensional
  sense is not guaranteed from the outset. They obtained this result
  keeping the finite lattice spacing for the extra direction finite,
  since it was still unclear how the low energy limit can be obtained.
  So there may exist the possibility that this mass-like term turns to
  vanish in such a limit.

   In this paper we do not try to overcome the first two problems.
  Rather we address the third question, especially the problem of
  the mass-like term. In fact, this is also not clear even in the
  continuum theory. So our question here is whether such a mass-like
  term found by Aoki and Hirose, arises or not, even in the low
  energy limit of the continuum theory. If this term does not appear
  in this case, the lattice regularization accounts for such breaking
  of the two-dimensional gauge invariance. Therefore we re-examine the
  system considered by Callan and Harvey \IE\ a continuum analog of
  Kaplan's formulation; in fact, using the explicit fermion
  propagator coupled to the domain-wall mass, we will evaluate the
  one-loop effective action for the domain-wall fermions.

   To perform the calculation in the continuum theory, we adopt the
  dimensional regularization slightly different from the ordinary ones
  in the treatment of the Dirac gamma matrices as follows:
  to the two-dimensional space parallel to the domain wall,
  we apply the usual dimensional regularization, and for the remaining
  direction parametrized by $\sigma$, the Dirac gamma matrix is held
  to be $\gamma^{2}=$ $i\gamma^{5}=$ $i\gamma^{0}\gamma^{1}$. Once we
  perform loop-integrations, we can find results finite.
  We can remove the regularization \IE\ $D \rightarrow 2$.
  Then we define the low energy limit as letting the domain-wall
  height $M$ go to infinity, {\it the large mass limit}.
  In this way, we show that the effective action in the anomaly-free
  case turns out to be gauge invariant as a two-dimensional theory;
  there is no mass-like term for the gauge boson in the large mass limit.

   This paper is organized as follows:
  In section $2$, we present all solutions to the Dirac equation
  coupled to the domain-wall mass including both the chiral modes and
  the massive modes. Then, adopting the canonical quantization, we derive
  the domain-wall fermion propagator. In section $3$, we check that the
  above-mentioned dimensional regularization respects the three-dimensional
  gauge invariance. Chandrasekharan
\ref\CHAN{S. Chandrasekharan, CU-TP-615, Columbia University preprint,
                                              October 1993. hep-th/9311050.}
  recently calculated a similar propagator in Euclidean space and
  showed the anomaly cancellation explicitly. We can see from the
  Lagrangian extended to extra dimensions that this regularization
  keeps the gauge invariance. But the above-mentioned propagator fails
  to be the kernel of the domain-wall Dirac equation under this
  regularization. So this seems not so obvious.
  Our main result is contained in section $4$, where we
  present the one-loop effective action for the domain-wall fermion in the
  large mass limit explicitly.
  Our conclusion and discussion are given in section $5$.
  In Appendix A, we present the properties of the functions appearing in the
  domain-wall fermion propagator. For the convenience in the perturbation
  calculation, we define a set of functions and give their large mass
  limit in Appendix B. In Appendix C, the calculation in section $4$
  is explained in some detail.

\newsec{Propagator for Domain-Wall Fermions}

   Following Kaplan \kaplan, we consider a Dirac fermion with a
  domain-wall mass
\eqn\A{
       \Ms = M \es ,
}
    where
\eqn\B{
        \qquad     \es = \cases{ 1   &    $(\sigma > 0)$
\cr
                                -1   &    $(\sigma < 0)$},
}
  in the three-dimensional Minkowsky space parametrized by
  $(x^0,x^1, x^2) = (x^0, x^1, \sigma)$, with metric
  $\eta_{IJ} = {\rm diag}.(1,-1,-1) \quad (I,J = 0,1,2)$.
  The Dirac equation is given by
\eqn\dirac{
\eqalign{
  0 & = \left[i \gamma^{I}\partial_{I} - \Ms \right]\psi(x,\sigma)
\cr
    &= \left[i \gamma^{\mu}\partial_{\mu} -
                 \left(\gamma^{5}\pder{\sigma} + \Ms \right) \right]
                                                               \psi(x,\sigma).
\cr
    & \qquad (I = 0,1,2;\quad  \mu = 0,1)
\cr}
}
  Note that if $\psi(x,\sigma)$ is a solution of \dirac ,
  then $\gamma^{5}\psi(x,-\sigma)$ is, too. Therefore we can decompose
  the solutions into odd/even eigenstates under the operation
\eqn\C{
  \psi(x,\sigma) \rightarrow \psi'(x,\sigma) = \gamma^{5}\psi(x,-\sigma),
}
  as
\eqnn\D
$$
\eqalignno{
  \psi^L(x,\sigma) &= {1 \over 2}\left(\psi(x,\sigma)
                                          - \gamma^{5}\psi(x,-\sigma) \right)
\cr
                   &= - \gamma^{5}\psi^L(x,-\sigma),
&\D\cr}
$$
  which we call a left-handed eigenstate, and
\eqn\E{
\eqalign{
  \psi^R(x,\sigma) &= {1 \over 2}\left(\psi(x,\sigma)
                                          + \gamma^{5}\psi(x,-\sigma) \right)
\cr
                   &= + \gamma^{5}\psi^R(x,-\sigma).
\cr}
}
  which we call a right-handed eigenstate. They have the same name of
  chiralities on the domain wall \IE\ at $\sigma = 0$,
\eqn\I{
  \gamma^{5}\psi^{{R\choose L}}(x,\sigma=0)
                                     = \pm\psi^{{R\choose L}}(x,\sigma=0).
}

   Now we present all the normalizable solutions of $\dirac$. We
  discard solutions that exponentially grow up as
  $\sigma \rightarrow \pm\infty$.
  The chiral zero modes which we expect to represent a two-dimensional
  chiral fermion, and which we hereafter call right-handed Weyl type
  wave functions, are given as
\eqn\F{
  U^R_W(p ;\sigma) = \sqrt{2M}e^{-M|\sigma|}{-i\sqrt{p_{0}} \choose 0}
                                                                    \theta(-p)
}
  which correspond to the positive-energy solutions, and
\eqn\G{
  V^R_W(p ;\sigma) = \sqrt{2M}e^{-M|\sigma|}{+i\sqrt{p_{0}} \choose 0}
                                                                    \theta(-p)
}
  which correspond to the negative-energy solutions,
  where $p_{0} = |p|$, and $p$ denotes the momentum $p_{1}$ conjugate to the
  coordinate $x^{1}$. By the exponential factor, these chiral
  modes are seen to be bound to the domain wall. Note that these
  solutions are only right-handed eigenstates and have the ordinary
  right-handed chirality at any value of $\sigma$. Such chirality
  for the chiral zero mode is determined by the sign
  of the domain-wall mass.

   On the other hand, there exist both right- and left-handed
  eigenstates in the massive modes which we call Dirac type wave
  functions. Such right-handed eigenstates are given as
\eqn\H{
  U^R_D(p,\omega;\sigma) = {1 \over \sqrt{p_{+}}}
                             {\omega\cos\omega\sigma
                                             - M(\sigma)\sin\omega\sigma
                          \choose
                                   -ip_{+}\sin\omega\sigma}
}
  with positive energy, and
\eqn\J{
  V^R_D(p,\omega;\sigma) = {1 \over \sqrt{p_{+}}}
                             {\omega\cos\omega\sigma
                                             - M(\sigma)\sin\omega\sigma
                           \choose
                                    +ip_{+}\sin\omega\sigma}
}
  with negative energy. While the left-handed eigenstates are given as
\eqn\K{
  U^L_D(p,\omega;\sigma) = {1 \over \sqrt{p_{-}}}
                             {     -ip_{-}\sin\omega\sigma
                          \choose
                                  \omega\cos\omega\sigma
                                             + M(\sigma)\sin\omega\sigma}
}
  with positive energy, and
\eqn\L{
  V^L_D(p,\omega;\sigma) = {1 \over \sqrt{p_{-}}}
                             {+ip_{-}\sin\omega\sigma
                          \choose
                                 \omega\cos\omega\sigma
                                            + M(\sigma)\sin\omega\sigma}
}
  with negative energy. Here
\eqn\M{
  p_{0} = E = \sqrt{p^2 + \omega^2 + M^2},
\qquad
  ( p = p_{1} )
}
  and $p_{\pm} = p_{0} \pm p_{1} = E \pm p. $
  $\omega$ denotes the momentum $p_{2}$ conjugate to the extra
  coordinate $\sigma$. Note that all these solutions are odd functions
  with respect to $\omega$, so that the solutions with $\pm\omega$ are
  not independent of each other.

  These wave functions are normalized such that
\eqnn\NORM \eqnn\NORMW
$$
\eqalignno{
                & \int d\sigma
                          U^{A \dag}_D(p,\omega';\sigma)U^B_D(p,\omega;\sigma)
\cr
              = & \int d\sigma
                          V^{A \dag}_D(p,\omega';\sigma)V^B_D(p,\omega;\sigma)
\cr
              = & 2E \times  {1 \over 2}(2\pi)\delta^{AB}
                  \left( \delta(\omega-\omega')-\delta(\omega+\omega') \right),
   \qquad                                                    ( A,B = L,R )
&\NORM\cr
                 &  \int d\sigma  U^{R \dag}_W(p;\sigma)U^R_W(p;\sigma)
\cr
               = & \int d\sigma  V^{R \dag}_W(p;\sigma)V^R_W(p;\sigma)
\cr
               = &  2|p|,
&\NORMW\cr}
$$
  and otherwise vanish. And they satisfy the following completeness
  relation:
\eqn\COMPLEW{
  {1 \over 2p_{0}}\left[ U^R_W(p,\sigma')U^{R \dag}_W(p,\sigma)
                               + V^R_W(p,\sigma')V^{R \dag}_W(p,\sigma) \right]
                  = Me^{-M(|\sigma'|+|\sigma|)}\proj{R}.
}
\eqn\COMPLED{
\eqalign{
   &\int^{\infty}_{-\infty} {d\omega \over (2\pi)}{1 \over 2p_{0}}\sum_{A=R,L}
                    \left[ U^A_D(p,\omega;\sigma')U^{A \dag}_D(p,\omega;\sigma)
                      + V^A_D(-p,\omega;\sigma')V^{A \dag}_D(-p,\omega;\sigma)
                                                                        \right]
\cr
      & \qquad  = \delta(\sigma'-\sigma) - Me^{-M(|\sigma|+|\sigma'|)}\proj{R},
\cr}
}
  with $\displaystyle \proj{{R \choose L}} = {{1 \pm \gamma^{5}}\over2}$,

   Now we are ready to derive the domain-wall fermion propagator
  following the cannonical quantization. The field operator
  $\psi(x,\sigma)$ is expanded as follows:
\eqn\PSIop{
  \psi(x,\sigma)
               = \psi^R_W(x,\sigma) + \psi^R_D(x,\sigma) + \psi^L_D(x,\sigma),
}
   where
\eqn\PSIW{
   \psi^R_W(x,\sigma) = \int^{\infty}_{-\infty}
                             {dp \over \sqrt{2\pi}\sqrt{2|p|}}
                         \left[ b(p)e^{-ip_\mu x^\mu}U^R_W(p,\sigma)
                          + d^{\dag}(p)e^{ip_\mu x^\mu}V^R_W(p,\sigma) \right],
}
  which consists of the chiral zero modes,
\eqn\PSIR{
\eqalign{
   \psi^R_D&(x,\sigma)
\cr
               = &\int^{\infty}_{-\infty}{dpd\omega \over (2\pi)\sqrt{2E}}
                  \left[
                    B_R(p,\omega)e^{-ip_\mu x^\mu}U^R_D(p,\omega;\sigma)
                                          +
                    D^{\dag}_R(p,\omega)e^{ip_\mu x^\mu}V^R_D(p,\omega;\sigma)
                                                                       \right],
\cr}
}
  which consists of the massive right-handed modes, and
\eqn\PSIL{
\eqalign{
   &\psi^L_D(x,\sigma)
\cr
              & = \int^{\infty}_{-\infty}{dpd\omega \over (2\pi)\sqrt{2E}}
                  \left[
                    B_L(p,\omega)e^{-ip_\mu x^\mu}U^L_D(p,\omega;\sigma)
                                          +
                    D^{\dag}_L(p,\omega)e^{ip_\mu x^\mu}V^L_D(p,\omega;\sigma)
                                                                       \right],
\cr}
}
  which consists of the massive left-handed modes. Since the wave functions
  $U_{D}$ and $V_{D}$ are odd under $\omega\leftrightarrow-\omega$,
  the annihilation (creation) operators
  $B^{(\dag)}$ and $D^{(\dag)}$ defined here satisfy the following
  relations:
\eqnn\N \eqnn\O
$$
\eqalignno{
  B^{(\dag)}_{{R \choose L}}(p,\omega)
                                  &= - B^{(\dag)}_{{R \choose L}}(p,-\omega),
&\N\cr
  D^{(\dag)}_{{R \choose L}}(p,\omega)
                                  &= - D^{(\dag)}_{{R \choose L}}(p,-\omega).
&\O\cr}
$$
  Because of these dependence relations, the $\omega$-integrations in
  \PSIR\ and \PSIL\ seem to count double the independent modes. But
  the double counting is properly avoided by the factor ${1 \over 2}$
  put in the normalization condition \NORM.
  From the equal-time commutation relation
\eqn\P{
  \{ \psi(t,x^1,\sigma), \psi(t,x'^1,\sigma')\}
                                     = \delta(x^1-x'^1)\delta(\sigma-\sigma'),
}
  we can derive the domain-wall fermion propagator \CHAN, in a similar
  fashion to the ordinary fermion case:
\eqnn\Q
$$
\eqalignno{
  \bra \T\psi(x,\sigma)\bar\psi(x',\sigma')\ket &= S_F(x,\sigma ; x',\sigma')
\cr
                                               &= S_F^W(x,\sigma ; x',\sigma')
                                                + S_F^D(x,\sigma ; x',\sigma'),
&\Q\cr}
$$
  where
\eqnn\R
$$
\eqalignno{
  S_F^W(x,\sigma ; x',\sigma')
                   &= \bra \T\psi^R_W(x,\sigma){\bar\psi}^R_W(x',\sigma')\ket
\cr
                   &= Me^{-M(|\sigma|+|\sigma'|)}
                         \int {d^2p \over i(2\pi)^2}
                          {1 \over -p^2-i\epsilon}e^{-ip(x-x')}\proj{R}\not{p},
&\R\cr}
$$
  which represents the two-dimensional chiral fermion propagator
  localized on the domain wall and comes from only the chiral zero
  modes. On the other hand,
\eqnn\S
$$
\eqalignno{
  S_F^D(x,\sigma ; x',\sigma')
       &= \sum_{A,B=R,L}\bra \T\psi^A_D(x,\sigma){\bar\psi}^B_D(x',\sigma')\ket
\cr
        = \int{d^2pd\omega \over i(2\pi)^3}&
                     {1 \over M^2+\omega^2-p^2-i\epsilon}e^{-ip(x-x')}
\cr
           &\big[
                  \proj{R}\left(\xi(\omega ; \sigma',\sigma)
                           + \not{p}\varphi_{-}(\omega ; \sigma,\sigma')\right)
\cr
                  &+ \proj{L}\left(\xi(\omega ; \sigma,\sigma')
                           + \not{p}\varphi_{+}(\omega ; \sigma,\sigma')\right)
                                                                        \big],
&\S\cr}
$$
  which comes from the massive modes, where
\eqnn\Ua \eqnn\Ub \eqnn\Uc
$$
\eqalignno{
  \xi(\omega ; \sigma, \sigma') &\equiv f_{+}(\omega,\sigma)\sin\omega\sigma'
                                       - f_{-}(\omega,\sigma')\sin\omega\sigma,
&\Ua\cr
  \varphi_{\pm}(\omega ; \sigma,\sigma') &\equiv {1 \over \omega^2+M^2}
                                  f_{\pm}(\omega,\sigma)f_{\pm}(\omega,\sigma')
                                          + \sin\omega\sigma\sin\omega\sigma' ,
&\Ub\cr
  f_{\pm}(\omega,\sigma) \equiv
                        & \omega\cos\omega\sigma \pm M(\sigma)\sin\omega\sigma.
&\Uc\cr
}
$$
  The properties of these functions are shown in Appendix A.
  Note that this propagator expression is valid in {\it any} $(2n+1)$
  dimensions, if $p_{\mu}$ is understood to be $2n$-dimensional one.

\newsec{Anomaly Cancellation}

   We couple an external Abelian gauge field to this system and
  investigate the gauge anomaly. Of course, it is obvious that there
  is no gauge anomaly in odd dimensions, because in such dimensions,
  all fermions are Dirac fermions so that gauge anomalies are
  automatically cancelled. Here we are not interested in whether
  this anomaly is cancelled or not, but in how this cancellation
  occurs. Since we have the right-handed chiral modes on
  the domain wall, it seems at first sight that this mode produces the
  gauge anomaly. Callan and Harvey\CalHav\ resolved this seemingly paradox.
  The gauge anomaly produced by this chiral mode is compensated with
  the current flow of the massive modes and the gauge invariance is
  maintained. Their argument is qualitatively clear and elegant,
  without using the explicit propagator which we derived as above.
  But for our purpose, we need to investigate the behavior of the
  massive modes explicitly, and make it clear whether this theory is
  chiral or not in the large mass limit $M \rightarrow \infty$.

   In calculating this gauge anomaly, we adopt the dimensional
  regularization in such a manner as explained in the introduction;
  namely, we use the usual dimensional regularization only for the
  first two dimensions and keep the $\sigma$ dimension intact.

   In this regularization scheme, our propagator is not the kernel of
  the free Dirac equation
     $\left[i \gamma^{\alpha}\partial_{\alpha} -
           \left(\gamma^{5}\pder{\sigma} + \Ms \right) \right]$,
  where $\alpha = (\mu,j)$ and $`j\,$' denotes the extra components due to
  the dimensional regularization. The reason is that it includes
  the ``chiral'' projection operators $\proj{{R \choose L}}$
  with $\gamma^{5}$ fixed to be $\gamma^{0}\gamma^{1}$ and so a
  similar situation occurs to that of usual chiral fermions;
  not all the components of dimensionally extended $\gamma^{\alpha}$
  anti-commute with $\gamma^{5}$. For later convenience' sake,
  we Fourier-transform our propagator with respect to only the
  directions parallel to the domain wall,
\eqn\V{
  S_F(x,\sigma ; x',\sigma') = \int{d^Dp \over i(2\pi)^D}
                                          e^{-ip(x-x')} S_F(p;\sigma,\sigma').
}
  Then the deviation from the Dirac equation can be verified as that
\eqn\W{
  \left[\dslash{p}+i\gamma^2\pder{\sigma}-M(\sigma)\right]
                      S_F(p;\sigma,\sigma')
    = -\delta(\sigma-\sigma')+\gamma^5\uslash{p}\Delta_F(p;\sigma,\sigma'),
}
  and
\eqn\X{
  S_F(p;\sigma',\sigma)
         \left[\dslash{p}
                   - i\gamma^2\overleftarrow{\pder{\sigma}}-M(\sigma)\right]
    = -\delta(\sigma-\sigma')+2\uslash{p}\bar\Delta_F(p;\sigma',\sigma),
}
  where $\dslash{p} = \gamma^{\alpha}p_{\alpha} = \lslash{p} + \uslash{p}$
  with $\lslash{p} = \gamma^{\mu}p_{\mu}$ and $\uslash{p} = \gamma^{j}p_{j}$,
  and
\eqn\Y{
  \Delta_F(p;\sigma,\sigma') =
                  \Delta^W_F(p;\sigma,\sigma') + \Delta^D_F(p;\sigma,\sigma'),
}
\eqn\Z{
\cases{
  \Delta^W_F(p;\sigma,\sigma') = Me^{-M(|\sigma|+|\sigma'|)}
                                      {1 \over -p^2-i\epsilon}\dslash{p},
\cr
\eqalign{
  \Delta^D_F(p;\sigma,\sigma')
                              = \int{d\omega \over 2\pi}
                                  {1 \over M^2+\omega^2-p^2-i\epsilon}
                                   \big[&\big(
                                      \xi(\omega;\sigma',\sigma) -
                                                  \xi(\omega;\sigma,\sigma')
                                                                        \big)
\cr
                                & +
                                  \dslash{p}\big(
                                        \varphi_{-}(\omega;\sigma,\sigma') -
                                          \varphi_{+}(\omega;\sigma,\sigma')
                                                               \big)\big],
\cr}
}
}
and
\eqn\1{
  \bar\Delta_F(p;\sigma',\sigma) =
          \bar\Delta^W_F(p;\sigma',\sigma) + \bar\Delta^D_F(p;\sigma',\sigma),
}
\eqn\2{
\cases{
  \bar\Delta^W_F(p;\sigma',\sigma) = -Me^{-M(|\sigma|+|\sigma'|)}
                                      {1 \over -p^2-i\epsilon}
                                                M(\sigma)\proj{R},
\cr
\eqalign{
  \bar\Delta^D_F&(p;\sigma',\sigma)
\cr
&
\eqalign{
                   =   \int{d\omega \over 2\pi}
                           {1 \over M^2+\omega^2-p^2-i\epsilon}
                            \big[& \proj{R}\big(
                                 \xi(\omega;\sigma,\sigma')
                                 - M(\sigma)\varphi_{-}(\omega;\sigma,\sigma')
                                                                          \big)
\cr
                             & +   \proj{L}\big(
                                 \xi(\omega;\sigma',\sigma)
                                  - M(\sigma)\varphi_{+}(\omega;\sigma,\sigma')
                                                                    \big)\big].
\cr}
\cr}
}}

   This situation is different from that for the ordinary
  three-dimensional Dirac fermions, which does not have such deviation.
  This fact makes the calculation for the anomaly cancellation
  nontrivial.

   We proceed to the calculation for the gauge anomaly. The gauge
  current is defined, as usual;
\eqn\3{
   J^{I}(x,\sigma) = \bar\psi(x,\sigma)\gamma^{I}\psi(x,\sigma).
}
  Then the total divergence of this current turns out to be
\eqn\anomaly{
\eqalign{
  \partial_{I}\langle&J^{I}(x,\sigma)\rangle
\cr
   = &ie  \int d\sigma'\int{d^2p \over (2\pi)^2}e^{-ipx}A_{I}(p,\sigma')
                \int{d^Dk \over i(2\pi)^D}
\cr
        &
 \eqalign{
         \times\Bigg[
        &\tr\left[\big\{(\lslash{p}+\dslash{k})+
                                 i\gamma^2\pder{\sigma}-M(\sigma)\big\}
                   S_F(p+k;\sigma,\sigma')\gamma^{I}
                                                  S_F(k;\sigma',\sigma)\right]
 \cr
       &- \tr\bigg[S_F(p+k;\sigma,\sigma')\gamma^{I}
                                                       S_F(k;\sigma',\sigma)
                 \big\{\dslash{k}
                   - i\gamma^2\overleftarrow{\pder{\sigma}}-M(\sigma)\big\}
                                                                 \bigg]\Bigg]
 \cr
 }
\cr
   = ie & \int d\sigma' \int{d^2p \over (2\pi)^2}e^{-ipx}A_{I}(p,\sigma')
\cr
        &\
 \eqalign{
  \times\int{d^Dk \over (2\pi)^Di}
        \Bigg[
       &\tr\left[\gamma^5\uslash{k}\Delta_F(p+k;\sigma,\sigma')
                               \gamma^{I}S_F(k;\sigma',\sigma) \right]
 \cr
     &- \tr\left[S_F(p+k;\sigma,\sigma')\gamma^{I}
                          2\uslash{k}\bar{\Delta}_F(k;\sigma',\sigma) \right]
                                                                        \Bigg],
 \cr
  }
\cr}
}
  where $A_{I}(p,\sigma)$ $(I=0,1,2)$ is the external
  Abelian gauge field with the coordinates $x^{\mu}$
  Fourier-transformed to the momenta $p^{\mu}$ and $e$ is the
  coupling constant.

   In fact, this anomaly  $\partial_{I}\langle J^{I}(x,\sigma)\rangle$
  vanishes, since the massive modes give the same contribution with
  the opposite sign as the chiral zero mode which has the exponential
  damping factor. In order to see this explicitly, firstly, we
  evaluate the first term in the square bracket of \anomaly .
  Then, we distinguish the contribution to this anomaly into four
  types, as follows:
\eqnn\4
$$
\eqalignno{
   \int{d^Dk \over (2\pi)^Di}
        & \tr\left[\gamma^5\uslash{k}\Delta_F(p+k;\sigma,\sigma')
                               \gamma^{I}S_F(k;\sigma',\sigma) \right]
\cr
    =  \int{d^Dk \over (2\pi)^Di} \bigg[
    \quad & \tr\left[\gamma^5\uslash{k}\Delta^W_F(p+k;\sigma,\sigma')
                               \gamma^{I}S^W_F(k;\sigma',\sigma) \right]
\cr
    +  &   \tr\left[\gamma^5\uslash{k}\Delta^W_F(p+k;\sigma,\sigma')
                               \gamma^{I}S^D_F(k;\sigma',\sigma) \right]
\cr
    +  &   \tr\left[\gamma^5\uslash{k}\Delta^D_F(p+k;\sigma,\sigma')
                               \gamma^{I}S^W_F(k;\sigma',\sigma) \right]
\cr
    +  &   \tr\left[\gamma^5\uslash{k}\Delta^D_F(p+k;\sigma,\sigma')
                               \gamma^{I}S^D_F(k;\sigma',\sigma) \right]
                                                                    \bigg].
&\4\cr}
$$
  They are evaluated to be
\eqnn\WW \eqnn\WD \eqnn\DW \eqnn\DD
$$
\eqalignno{
  \int&{d^Dk \over (2\pi)^Di}
    \tr\left[\gamma^5\uslash{k}\Delta^W_F(p+k;\sigma,\sigma')
                            \gamma^{I}S^W_F(k;\sigma',\sigma) \right]
\cr
              &\qquad = -{1 \over 4\pi} M^2 e^{-2M\absig}
                                  [\epsilon^{2 \mu I}+g^{\mu I}]p_{\mu}.
&\WW\cr
  \int&{d^Dk \over (2\pi)^Di}
    \tr\left[\gamma^5\uslash{k}\Delta^W_F(p+k;\sigma,\sigma')
                            \gamma^{I}S^D_F(k;\sigma',\sigma) \right]
\cr
&
\eqalign{
               \qquad = & - {1 \over 4\pi} M e^{-M\absig }
\cr
                   & \qquad\times[\epsilon^{2 \mu I}
                          (2\delta(\sigma-\sigma')-Me^{-M\absig})-g^{\mu I}
                                                        M e^{-M\absig}]p_{\mu}.
\cr}
&\WD\cr
  \int&{d^Dk \over (2\pi)^Di}
    \tr\left[\gamma^5\uslash{k}\Delta^D_F(p+k;\sigma,\sigma')
                            \gamma^{I}S^W_F(k;\sigma',\sigma) \right]
\cr
              &\qquad =  {1 \over 4\pi} M^2 e^{-2M\absig}
                                  [\epsilon^{2 \mu I}+g^{\mu I}]p_{\mu}.
&\DW\cr
  \int&{d^Dk \over (2\pi)^Di}
    \tr\left[\gamma^5\uslash{k}\Delta^D_F(p+k;\sigma,\sigma')
                               \gamma^{I}S^D_F(k;\sigma',\sigma) \right]
\cr
&
\eqalign{
               \qquad =   & {1 \over 4\pi}  M e^{-M\absig}
\cr
                   & \qquad\times[\epsilon^{2 \mu I}
                          (2\delta(\sigma-\sigma')-Me^{-M\absig})-g^{\mu I}
                                                        M e^{-M\absig}]p_{\mu}.
\cr}
&\DD
\cr}
$$
  From these equations, we see that \WW\ and \DW\ cancel each other,
  and so do \WD\  and \DD\ . As is different from the ordinary Dirac
  fermion, the contribution \DD\ has the exponential damping factor
  and is of chiral type, despite that it comes from  massive
  Dirac-type modes alone. We see that
  similar things also happen for the other remaining terms in the square
  bracket of \anomaly . Thus, the gauge anomaly
  $ \partial_{I}\langle J^{I}(x,\sigma)\rangle$ turns out to be
  zero \CHAN.

\eqn\5{
  \partial_{I}\langle J^{I}(x,\sigma)\rangle = 0.
}

\newsec{One-Loop Effective Action}

   Usually, if we adopt the ordinary dimensional regularization
  for chiral fermions, we can not get a gauge invariant answer
  even in anomaly-free chiral gauge theories. The reason is
  well known to be that such a regularization does not respect
  the gauge invariance. On the other hand in the case under
  consideration, the contribution from the chiral modes are summed
  up with that from the massive modes to yield a three-dimensional
  gauge invariant result, as we have seen in the previous section.
  What happens there is the following; the functions
  $\varphi_{\pm}(\omega ; \sigma,\sigma')$ included in the Dirac type
  propagator represent ``the densities of the massive modes''
  for the right- and left-handed eigenstate, respectively.
  Then the difference between $\varphi_{\pm}(\omega;\sigma,\sigma')$'s
  accounts for the mismatching of the number of modes between right-
  and left-handed ones. Thus we may expect that the contribution
  from the massive modes can be essentially decomposed into two parts:
  one from a massive Dirac fermion and the other from the just-mentioned
  difference; that is, in the large mass limit $M \rightarrow \infty$,
  only the former decouples, while the difference survives the limit
  and is summed up with the chiral mode contribution to yield a gauge
  invariant answer even in our dimensional regularization.
  We show in this section that this is the case in the present model,
  evaluating the one-loop effective action in the gauge $A_2 = 0$
  as follows:
\eqn\6{
  e^{i\Gamma[A]} \equiv
   {\rm Det}\left[i\dslash{\partial}-M(\sigma)+e\lslash{A}(x,\sigma)\right].
}
\eqn\7{
  \Gamma[A] = -{e^2 \over 2}\int{d^2p \over (2\pi)^2}d\sigma d\sigma'
                 A_{\mu}(-p,\sigma)\Pi^{\mu\nu}(p;\sigma,\sigma')
                                 A_{\nu}(p,\sigma') + \cdot \cdot \cdot
}
  with
\eqn\8{
  \Pi^{\mu\nu}(p;\sigma,\sigma') \equiv  \int{d^Dk \over i(2\pi)^D}
                \tr\left[\gamma^{\mu}S_F(p+k;\sigma,\sigma')
                                   \gamma^{\nu}S_F(k;\sigma',\sigma)\right],
}
  where the dots denotes the higher order terms in $A_{\mu}$.
  The vacuum polarization $\Pi^{\mu\nu}(p;\sigma,\sigma')$ has
  three distinct contributions depending on whether the two internal
  fermion lines are Weyl-type $S^W_F$ or Dirac-type $S^D_F$:
\eqnn\8{
$$
\eqalignno{
  \Pi^{\mu\nu}_W(p;\sigma,\sigma')
   &= \int{d^Dk \over (2\pi)^Di}\tr\left[\gamma^{\mu}S^W_F(p+k;\sigma,\sigma')
                                   \gamma^{\nu}S^W_F(k;\sigma',\sigma)\right],
\cr
  \Pi^{\mu\nu}_M(p;\sigma,\sigma')
   &= \int{d^Dk \over (2\pi)^Di}\tr\left[\gamma^{\mu}S^W_F(p+k;\sigma,\sigma')
                                   \gamma^{\nu}S^D_F(k;\sigma',\sigma)\right],
\cr
  \Pi^{\mu\nu}_D(p;\sigma,\sigma')
   &= \int{d^Dk \over (2\pi)^Di}\tr\left[\gamma^{\mu}S^D_F(p+k;\sigma,\sigma')
                                   \gamma^{\nu}S^D_F(k;\sigma',\sigma)\right].
&\8\cr}
$$

   We calculate these contributions in our regularization scheme. Since
  these quantities turn out to be finite, we can remove the
  regularization, \IE\ $D \rightarrow 2$. After that, we take the
  magnitude $M$ of the domain-wall mass to infinity.

   As we mentioned above, the ordinary dimensional regularization breaks
  the gauge invariance in the chiral gauge theories, so we cannot get
  even the parity-even contribution of e.g., the vacuum polarization,
  in a gauge invariant way. This is the case also here, but only
  in the pure Weyl-contribution $\Pi^{\mu\nu}_W(p;\sigma,\sigma')$,
  which is expected to represent the vacuum polarization of the
  two-dimensional chiral gauge theory. However we will see later that
  the large mass limit of the total vacuum polarization
  $\Pi^{\mu\nu}(p;\sigma,\sigma')$ becomes gauge invariant in the
  two-dimensional sense, if we consider anomaly-free chiral gauge theories.

   In order to see this, we would like to show how these contributions
  are summed up to yield a gauge invariant result. So we evaluate
  each of the contributions separately now.

\subsec{Pure Weyl-type contribution $\Pi^{\mu\nu}_W(p;\sigma,\sigma')$}

   Removing the dimensional regularization, the pure Weyl-type contribution
  $\Pi^{\mu\nu}_W(p;\sigma,\sigma')$ is seen to be
\eqn\weylpi{
\eqalign{
  \Pi^{\mu\nu}_W(p;\sigma,\sigma')
   =&{1 \over 4\pi}M^2e^{-2\absig}{1 \over p^2}\ptensor{L}p_{\rho}p_{\sigma}
\cr
   \sitarel{\rightarrow}{M \rightarrow \infty}&({1 \over 4\pi}){1 \over p^2}
       \left[ 2(p^{\mu}p^{\nu}-g^{\mu\nu}p^2) -
         (p^{\mu}\epsilon^{\nu\rho}+p^{\nu}\epsilon^{\mu\rho})p_{\rho}\right]
                                                \delta(\sigma)\delta(\sigma')
\cr
     &+ {1 \over 4\pi}\left(g^{\mu\nu}+\epsilon^{\mu\nu}\right)
                                                \delta(\sigma)\delta(\sigma'),
\cr}
}
  where
\eqn\9{
  \ptensor{{L \choose R}} = \tr[\proj{{L \choose R}}\gamma^{\mu}
                                 \gamma^{\rho}\gamma^{\nu}\gamma^{\sigma}],
}
  and in the large mass limit $M \rightarrow \infty$,
\eqn\AA{
  \lim_{M \rightarrow \infty}Me^{-2M|\sigma|} = \delta(\sigma).
}

   From \weylpi, we can see that even the parity-even part of this quantity
  is not gauge invariant.

\subsec{Mixed-type contribution $\Pi^{\mu\nu}_M(p;\sigma,\sigma')
                                      + \Pi^{\nu\mu}_M(-p;\sigma',\sigma)$}

   We have only to evaluate the part $\Pi^{\mu\nu}_M(p;\sigma,\sigma')$
  of the mixed-type contribution. After the loop-integration over the
  momentum $k$, we can remove the regularization, as mentioned above.
  Since this quantity has no infra-red singularity, we can
  Taylor-expand it with respect to the momentum $p^{\mu}$. We can see
  from the dimensional analysis that all the terms other than
  the leading order one in this expansion, turn to be
  zero in the large mass limit $M \rightarrow \infty$.
\eqn\mixedpi{
\eqalign{
  \Pi^{\mu\nu}_M(p;\sigma,\sigma')
   = -({1 \over 4\pi})&Me^{-2M|\sigma|}(g^{\mu\nu}-\epsilon^{\mu\nu})
                                                   \delta(\sigma-\sigma')
\cr
     -({1 \over 4\pi})&Me^{-2M\absig}\ptensor{L}p_{\rho}p_{\sigma}
\cr
       &\times\int^{1}_{0}d\alpha(1-\alpha)\int{d\omega \over 2\pi}
         {\varphi_{-}(\omega;\sigma,\sigma') \over
                                   \omega^2+M^2-(1-\alpha)p^2}
\cr
     + O({p^2 \over M}&).
\cr}
}
  with $\alpha$ being Feynman parameter. Since the second term in
  $\mixedpi$ is also seen to vanish in the limit
  $M \rightarrow \infty$,
\eqn\AB{
  Me^{-2M\absig}\int^{1}_{0}d\alpha(1-\alpha)\int{d\omega \over 2\pi}
         {\varphi_{-}(\omega;\sigma,\sigma') \over
                                   \omega^2+M^2}
              \sitarel{\rightarrow}{M \rightarrow \infty} 0,
}
  we obtain
\eqn\AC{
 \Pi^{\mu\nu}_M(p;\sigma,\sigma')
   \sitarel{\rightarrow}{M \rightarrow \infty}
      -({1 \over 4\pi})(g^{\mu\nu}-\epsilon^{\mu\nu})
                                              \delta(\sigma)\delta(\sigma').
}
   Therefore the mixed-type contribution turns out to be
\eqn\AD{
  \Pi^{\mu\nu}_M(p;\sigma,\sigma')+\Pi^{\nu\mu}_M(-p;\sigma',\sigma)
     \sitarel{\rightarrow}{M \rightarrow \infty}
        (-2) \times {1 \over 4\pi}g^{\mu\nu}\delta(\sigma)\delta(\sigma').
}

\subsec{Pure Dirac-type contribution $\Pi^{\mu\nu}_D(p;\sigma,\sigma')$}

   The calculation for this pure Dirac-type contribution
  $\Pi^{\mu\nu}_D(p;\sigma,\sigma')$ is rather involved. Here we
  only present the result in the large mass limit. The derivation are
  given in detail in Appendix C.
\eqn\DIP{
\eqalign{
  \Pi^{\mu\nu}_D(p;\sigma,&\sigma')
\cr
     = &{\Gamma(2-{D\over 2})\over(4\pi)^{{D \over  2}}}\int^{1}_{0}d\alpha
        \int{d\omega d\omega' \over (2\pi)^2}
\cr
&
\eqalign{
          \times\Bigg[
            &{1 \over [M^2+\alpha\omega^2+(1-\alpha){\omega'}^2
                                 -\alpha(1-\alpha)p^2]^{1-{D\over2}}}
\cr
            &\quad\times\Big\{\tr(\proj{L}\gamma^{\mu}\gamma^{\nu})
                         \big(\varphi_{+}(\omega;\sigma,\sigma')
                                    -\varphi_{-}(\omega;\sigma,\sigma')\big)
                                     \varphi_{+}(\omega';\sigma',\sigma)
\cr
            &\qquad+\tr(\proj{R}\gamma^{\mu}\gamma^{\nu})
                         \big(\varphi_{-}(\omega;\sigma,\sigma')
                                     -\varphi_{+}(\omega;\sigma,\sigma')\big)
                                     \varphi_{-}(\omega';\sigma',\sigma)\Big\}
\cr}
\cr
&
\eqalign{
            -&{1 \over [M^2+\alpha\omega^2+(1-\alpha){\omega'}^2
                                 -\alpha(1-\alpha)p^2]^{2-{D\over2}}}
\cr
          &\qquad\times\pder{\sigma}\Big\{M(\sigma)
              \big[\tr(\proj{L}\gamma^{\mu}\gamma^{\nu})
                        \varphi_{+}(\omega;\sigma,\sigma')
                                     \varphi_{+}(\omega';\sigma',\sigma)
\cr
                &\qquad\qquad\qquad-\tr(\proj{R}\gamma^{\mu}\gamma^{\nu})
                        \varphi_{-}(\omega;\sigma,\sigma')
                                     \varphi_{-}(\omega';\sigma',\sigma)\big]
                                                                 \Big\}\Bigg],
\cr}
\cr}
}
  where we neglect the irrelevant terms in the large mass limit.
  Using the formulae of the $\omega$-integration for $\varphi_{\pm}$
  in Appendix A, the first term are seen to be finite at $D=2$ and can
  be easily evaluated. As for the second term, the result are derived
  in Appendix C. As we can see there, this term is also finite.
  In the large mass limit, we obtain
\eqn\diracpi{
  \Pi^{\mu\nu}_D(p;\sigma,\sigma')
   \sitarel{\rightarrow}{M \rightarrow \infty}
         ({1 \over 4\pi})\left[
                    (g^{\mu\nu}-\epsilon^{\mu\nu})
                                             \delta(\sigma)\delta(\sigma')
                    +\epsilon^{\mu\nu}\pder{\sigma}
                              \big(\epsilon(\sigma)\delta(\sigma-\sigma')\big)
                                                                      \right].
}
\vskip .1in

\subsec{The total vacuum polarization and the one-loop effective action}

\vskip .1in

   Summing up three contributions evaluated in the previous subsections,
  we find the total vacuum polarization $\Pi^{\mu\nu}(p;\sigma,\sigma')$
  in the large mass limit to be
\eqnn\vacpol
$$
\eqalignno{
  \Pi^{\mu\nu}(p;\sigma,\sigma')
    &=\Pi^{\mu\nu}_W(p;\sigma,\sigma')
       +\Pi^{\mu\nu}_M(p;\sigma,\sigma')+\Pi^{\nu\mu}_M(-p;\sigma',\sigma)
       +\Pi^{\mu\nu}_D(p;\sigma,\sigma')
\cr
&
\eqalign{
    \sitarel{\rightarrow}{M \rightarrow \infty}\quad
        &{1 \over 2\pi}{1 \over p^2}(p^{\mu}p^{\nu}-g^{\mu\nu}p^2)
                                                 \delta(\sigma)\delta(\sigma')
\cr
        &-{1 \over 4\pi}{1 \over p^2}
              (p^{\mu}\epsilon^{\nu\rho}+p^{\nu}\epsilon^{\mu\rho})p_{\rho}
                                                  \delta(\sigma)\delta(\sigma')
\cr
        &+{1 \over 4\pi}\epsilon^{\mu\nu}
                 \pder{\sigma}\big(\epsilon(\sigma)\delta(\sigma-\sigma')\big).
\cr}
\cr}
$$

   Finally, when the magnitude $M$ of the domain-wall mass goes to infinity,
  the one-loop effective action $\Gamma[A]$ has a limit
\eqn\ACTION{
\eqalign{
  \Gamma[A]
      \sitarel{\rightarrow}{M \rightarrow \infty}
    -{e^2 \over 4\pi}\int {d^2p \over (2\pi)^2}
     \bigg[&{1 \over p^2} A_{\mu}(-p,\sigma=0)
                             (p^{\mu}p^{\nu}-g^{\mu\nu}p^2)A_{\nu}(p,\sigma=0)
\cr
         & +{1 \over p^2} p^{\rho}A_{\rho}(-p,\sigma=0)
                                   \epsilon^{\mu\nu}p_{\mu}A_{\nu}(p,\sigma=0)
\cr
         & +\int d\sigma{1 \over 2}\epsilon^{\mu\nu}
                       A_{\mu}(-p,\sigma)\epsilon(\sigma)
                                         \pder{\sigma}A_{\nu}(p,\sigma)\bigg]
\cr
         + \cdot\cdot\cdot.
\cr}
}

   We see that the first term, the parity-even part, is
  gauge invariant in two-dimensional sense, which is expected as
  the contribution in two-dimensional chiral gauge theories. However
  note that this part could not be made gauge invariant, if the
  contribution came from only the chiral zero mode. On the other hand,
  we can regard the second term as the chiral anomaly from
  a two-dimensional chiral fermion, while the third term represents the
  current flow or the Chern-Simons term, which may be compared with
  the Goldstone-Wilczek current discussed by Callan and Harvey \CalHav.

   If we consider anomaly-free chiral Schwinger models instead, the
  second and third terms in \ACTION\ are absent, and so a gauge
  invariant one-loop effective action results. It may be interesting
  to compare this result with the lattice version \AokHir\
  by Aoki and Hirose, though in the latter the contribution from
  the doublers might be important.

\newsec{Conclusions and Discussions}

   We analyzed the would-be chiral Schwinger models in the continuum
  version of the Kaplan's formulation \kaplan. At first we derived the
  domain-wall fermion propagator in the explicit form.
  It consists of two parts: the propagator of the chiral mode bound
  to the domain wall and the propagator of the massive modes.
  Using this propagator, we performed the perturbative expansion for
  the one-loop effective action of the domain-wall fermion.
  In this calculation, we adopt the dimensional regularization
  explained in the text, which respects the gauge invariance.
  In fact we have concretely shown how the divergence of the gauge
  current vanishes in this regularization. After we have verified that
  the super-renormalizability of this theory allows us to remove
  the regularization, we made the domain-wall height $M$ go to
  infinity in the effective action. Then we have shown that this
  action turns out to be gauge invariant in two-dimensional sense
  in the large mass limit, if we consider anomaly-free cases.
  This is an interesting result, because the three-dimensional
  gauge invariance reduce to the two-dimensional one in the low energy
  theory without adding any noninvariant counterterm; namely, the
  whole domain wall system serves as a gauge-invariant regularization
  for the two-dimensional chiral Schwinger models. On the other hand
  for anomalous case, we got both the anomaly term which is expected
  from the chiral zero mode and the Chern-Simons term from the massive
  modes as discussed by Callan and Harvey \CalHav, besides the above
  gauge invariant parity-even term.

   As we mentioned in the introduction, Aoki and Hirose \AokHir\ have
  calculated the one-loop effective action in the lattice counterpart of
  the models discussed here. Since the extra space remains discretized
  in their calculation, we cannot directly compare their results with ours.
  But our results indicate the possibility that there exists the
  lattice counterpart of the large mass limit we discussed here, and
  that the gauge boson mass-like term vanishes in such a limit.
  Otherwise, the nondecoupling effect of the doublers should account
  for the mass-like term. So it is an interesting attempt to identify
  such a limit in their effective action.

   If we can succeed in it, the problem of the mass-like term
  will be harmless in thinking of the gauge field dependent
  on the extra coordinate. Then the remaining problems are
  how to reduce the freedom of the $(2n+1)$-dimensional gauge bosons
  to $2n$-dimensional ones and how to separate the chiral zero-mode of the
  opposite chirality on the anti domain wall, as mentioned in the
  introduction. We surely need further consideration on these issues.

   So far, we have discussed as target theories the two-dimensional
  models which are super-renormalizable.
  If we consider the four-dimensional models,
  we encounter the UV divergence. Then we cannot avoid to modify
  the definition of the large mass limit. Therefore our results
  can not be extended to that case straightforwardly.
  We would like to discuss this problem elsewhere.

\vskip .4cm

\centerline{{\bf Acknowledgements}}

   We would like to thank T.~Kugo and S.~Aoki for valuable discussions.
  We are also grateful to T.~Kugo for careful reading of the manuscript.

\appendix{A}{ Properties of the functions
                             appearing in the propagator }
\medskip

   The functions
\eqnn\AEa \eqnn\AEb \eqnn\AEc
$$
\eqalignno{
  f_{\pm}(\omega,\sigma) \equiv & \omega\cos\omega\sigma
                                              \pm M(\sigma)\sin\omega\sigma,
&\AEa\cr
  \varphi_{\pm}(\omega ; \sigma,\sigma') &\equiv {1 \over \omega^2+M^2}
                               f_{\pm}(\omega,\sigma)f_{\pm}(\omega,\sigma')
                                          + \sin\omega\sigma\sin\omega\sigma' ,
&\AEb\cr
  \xi(\omega ; \sigma, \sigma') &\equiv f_{+}(\omega,\sigma)\sin\omega\sigma'
                                      - f_{-}(\omega,\sigma')\sin\omega\sigma
\cr
                                &= -\omega\sin\omega(\sigma-\sigma')
                                          + \left(M(\sigma)+M(\sigma')\right)
                                             \sin\omega\sigma\sin\omega\sigma'.
&\AEc\cr}
$$
   appearing in the propagator, satisfy the following differential equations:
\eqn\AF{
\eqalign{
          \left[\pder{\sigma'}+M(\sigma')\right]\xi(\omega ; \sigma,\sigma')
                   &= (\omega^2+M^2)\varphi_{+}(\omega ; \sigma,\sigma').
\cr
          \left[\pder{\sigma}-M(\sigma)\right]\xi(\omega ; \sigma,\sigma')
                   &= -(\omega^2+M^2)\varphi_{-}(\omega ; \sigma,\sigma').
\cr
}
}
\eqn\AG{
\eqalign{
  \left[\pder{\sigma}+M(\sigma)\right]\varphi_{-}(\omega ; \sigma,\sigma')
                   &= \xi(\omega ; \sigma,\sigma').
\cr
  \left[\pder{\sigma}-M(\sigma)\right]\varphi_{+}(\omega ; \sigma,\sigma')
                   &= -\xi(\omega ; \sigma',\sigma).
\cr
}
}

   Some useful formulae, in particular, for proving the completeness of the
  wave functions are:
\eqn\AH{
  \bullet \quad e^{\pm i\omega|\sigma|} = \cos\omega\sigma
                                            \pm
                                     i\epsilon(\sigma)\sin\omega\sigma.
}
\eqn\AJ{
  \bullet  \quad \cases{ \int{d\omega \over 2\pi}
                                          \varphi_{+}(\omega ; \sigma,\sigma')
                              = \delta(\sigma - \sigma')&
\cr
                   \int{d\omega \over 2\pi}\varphi_{-}(\omega ; \sigma,\sigma')
                             = \delta(\sigma - \sigma')
                                     - Me^{-M(|\sigma|+|\sigma'|)}&}
}

\appendix{B}{ Definition of the functions
                      appearing in the perturbation calculation
                             and their limiting form in $M\rightarrow\infty$}
\medskip

   When we calculate the one-loop effective action after integrating
  over the loop momentum parallel to the domain wall, we encounter
  the following function
\eqn\AK{
  I_n(\sigma ; a)
     \equiv
      M \int^{\infty}_{0}{\tau}^{n-1}
                    \exp\left(-\tau-{M^2 \over 4\tau}{\sigma}^2a\right)d\tau,
\qquad                               ( {\rm for}\ a > 0 )
}
  which may be compared with the modified Bessel function $K_{\nu}(x)$.

  In the limit $M\rightarrow\infty$,
\eqn\AL{
   I_n(\sigma ; a)  \rightarrow
            ({4\pi \over a})^{{1 \over 2}}\Gamma(n+{1 \over 2})\delta(\sigma).
}

  Another function we meet is:
\eqn\AM{
   J_n(\sigma,\sigma'; a)
     \equiv
      M^2 \int^{\infty}_{0}d\tau{\tau}^{n-1}
       \exp\left(-\tau-{M^2 \over 4\tau}
           \left\{{{\sigma}^2 \over a}+{{\sigma'}^2 \over 1-a}\right\}\right).
\qquad                               ( {\rm for}\ a > 0 )
}
  This has a limit,
\eqn\AN{
   J_n(\sigma,\sigma'; a)
    \sitarel{\rightarrow}{M \rightarrow \infty}
     4\pi\Gamma(n+1)(a(1-a))^{{1 \over 2}}\delta(\sigma)\delta(\sigma').
}

  In addition, the following relations hold:
\eqnn\AOa \eqnn\AOb
$$
\eqalignno{
  &\pder{\alpha}f^{1 \over 2}(\alpha)I_{n-1}(\sigma; f(\alpha))
    ={2\over M^2}\Big(\pder{\alpha}f^{-{1\over2}}(\alpha)\Big)
       \dpder{\sigma}I_{n}(\sigma; f(\alpha)),
&\AOa\cr
  &\pder{\alpha}
     \big({1 \over f(\alpha)(1-f(\alpha))}\big)^{1 \over 2}
                                      J_{n-1}(\sigma,\sigma'; f(\alpha))
\cr
    &={1\over M^2}
            \Big({1\over f(\alpha)(1-f(\alpha))}\Big)^{1 \over 2}
         \big(\pder{\alpha}f(\alpha)\big)
                \Big(\dpder{\sigma}-\dpder{\sigma'}\Big)
                                    J_{n}(\sigma,\sigma'; f(\alpha)),
&\AOb
\cr}
$$
  where $f(\alpha)$ is an arbitrary function of $\alpha$.

\appendix{C}{Calculation for the pure
                 Dirac-type contribution $\Pi^{\mu\nu}_D$}
\medskip

   In this appendix, we outline the derivation of \diracpi. The pure
  Dirac-type contribution is defined by

\eqn\EA{
  \Pi^{\mu\nu}_D(p;\sigma,\sigma')
    = \int{d^Dk \over (2\pi)^Di}\tr\left[\gamma^{\mu}S^D_F(p+k;\sigma,\sigma')
                                    \gamma^{\nu}S^D_F(k;\sigma',\sigma)\right].
}
  Performing the loop-integration, we obtain
\eqn\EB{
\eqalign{
  \Pi^{\mu\nu}_D(p;\sigma,\sigma')
      = &{\Gamma(2-{D\over 2})\over(4\pi)^{\DT}}
          \int^{1}_{0}d\alpha\int{d\omega d\omega' \over (2\pi)^2}
\cr
        &    \times [ \Pi^{\mu\nu}_a(p;\omega,\omega';\alpha)
                     +\Pi^{\mu\nu}_b(p;\omega,\omega';\alpha)
                     +\Pi^{\mu\nu}_c(p;\omega,\omega';\alpha)]
\cr}
}
  with the Feynman parameter $\alpha$, where
\eqnn\ECa \eqnn\ECb \eqnn\ECc
$$
\eqalignno{
&\eqalign{
  \Pi^{\mu\nu}_a(p;\omega,\omega';\alpha)
    = -&\DENNO{1-\DT}
\cr
     &\times \Big\{\tr[\proj{L}\gamma^{\mu}\gamma^{\nu}]
                        \varphi_{-}(\omega;\sigma,\sigma')
                                     \varphi_{+}(\omega';\sigma',\sigma)
\cr
               & \qquad+\tr[\proj{R}\gamma^{\mu}\gamma^{\nu}]
                        \varphi_{+}(\omega;\sigma,\sigma')
                                    \varphi_{-}(\omega';\sigma',\sigma)\Big\},
\cr}
&\ECa\cr
&\eqalign{
  \Pi^{\mu\nu}_b(p;\omega,\omega';\alpha)
    = -&{\alpha(1-\alpha) \over \DENO{2-\DT}}
\cr
        &\times \Big\{\tr[\proj{L}\gamma^{\mu}\lslash{p}\gamma^{\nu}\lslash{p}]
                        \varphi_{-}(\omega;\sigma,\sigma')
                                     \varphi_{-}(\omega';\sigma',\sigma)
\cr
             & \qquad+\tr[\proj{R}\gamma^{\mu}\lslash{p}\gamma^{\nu}\lslash{p}]
                        \varphi_{+}(\omega;\sigma,\sigma')
                                     \varphi_{+}(\omega';\sigma',\sigma)\Big\}
\cr}
&\ECb\cr
&\eqalign{
  \Pi^{\mu\nu}_c(p;\omega,\omega';\alpha)
    = &\ \DENNO{2-\DT}
\cr
                 &\times\Big\{\tr[\proj{L}\gamma^{\mu}\gamma^{\nu}]
                      \xi(\omega;\sigma',\sigma)\xi(\omega';\sigma',\sigma)
\cr
                 & \qquad+\tr[\proj{R}\gamma^{\mu}\gamma^{\nu}]
                      \xi(\omega;\sigma,\sigma')\xi(\omega';\sigma,\sigma')
                                                                      \Big\}.
\cr}
&\ECc
\cr}
$$

   From the differential equations in Appendix A, we can verify the
  following equations:
\eqn\derxi{
\eqalign{
\left\{\eqalign{
  &\xi(\omega;\sigma',\sigma)\xi(\omega';\sigma',\sigma)  \cr
  &\xi(\omega;\sigma,\sigma')\xi(\omega';\sigma,\sigma') \cr}\right\}
   &= {1 \over 2}\dpder{\sigma}\left(\varphi_{\pm}(\omega;\sigma,\sigma')
                                   \varphi_{\pm}(\omega';\sigma',\sigma)\right)
\cr
   &  \mp\pder{\sigma}\left(M(\sigma)\varphi_{\pm}(\omega;\sigma,\sigma')
                                 \varphi_{\pm}(\omega';\sigma',\sigma)\right)
\cr
   &  +{1 \over 2}(\omega^2+{\omega'}^2+2M^2)
                               \left(\varphi_{\pm}(\omega;\sigma,\sigma')
                                  \varphi_{\pm}(\omega';\sigma',\sigma)\right).
\cr}
}
  We substitute these equations into $\Pi^{\mu\nu}_c(p;\omega,\omega';\alpha)$.
  Then note that the third term in \derxi\ can be replaced by
  $({\omega'}^2+M^2)\varphi_{\pm}\varphi_{\pm}$ since the
  difference $\propto$ $(\omega^2-{\omega'}^2)$ vanishes
  in the integral \EB\foot{This can be seen by noting the even property of
  the multiplied factor in \EB\ under $\omega \leftrightarrow \omega'$
  and $\alpha \leftrightarrow 1-\alpha$.}\llap. Further we can rewrite
  it into
\eqn\ED{
\eqalign{
  &\DENOO
     \varphi_{\pm}(\omega;\sigma,\sigma')\varphi_{\pm}(\omega';\sigma',\sigma)
\cr
  & -\alpha\pder{\alpha}\DENOO
      \varphi_{\pm}(\omega;\sigma,\sigma')\varphi_{\pm}(\omega';\sigma',\sigma)
\cr
  & +\alpha^2p^2
     \varphi_{\pm}(\omega;\sigma,\sigma')\varphi_{\pm}(\omega';\sigma',\sigma),
\cr}
}
  the first term of which is combined with
  $\Pi^{\mu\nu}_a(p;\omega,\omega';\alpha)$ to yield
\eqn\EE{
\eqalign{
         &\DENNO{1-\DT}
\cr
            &\quad\times\Big\{\tr[\proj{L}\gamma^{\mu}\gamma^{\nu}]
                         \big(\varphi_{+}(\omega;\sigma,\sigma')
                                    -\varphi_{-}(\omega;\sigma,\sigma')\big)
                                     \varphi_{+}(\omega';\sigma',\sigma)
\cr
            &\qquad+\tr[\proj{R}\gamma^{\mu}\gamma^{\nu}]
                         \big(\varphi_{-}(\omega;\sigma,\sigma')
                                     -\varphi_{+}(\omega;\sigma,\sigma')\big)
                                    \varphi_{-}(\omega';\sigma',\sigma)\Big\}.
\cr}
}
  This is the first term in \DIP. Using the formulae for the
  $\omega$-integration of $\varphi_{\pm}$ in Appendix A, we can see
  that this part in \EB\ is finite as $D\rightarrow2$ and gives
\eqnn\EF
$$
\eqalignno{
&\eqalign{
{\Gamma(2-{D\over 2})\over(4\pi)^{\DT}}
     \int^{1}_{0}d\alpha\int{d\omega d\omega' \over (2\pi)^2}&\DENNO{1-\DT}
\cr
             \times\Big\{\tr[\proj{L}&\gamma^{\mu}\gamma^{\nu}]
                         \big(\varphi_{+}(\omega;\sigma,\sigma')
                                    -\varphi_{-}(\omega;\sigma,\sigma')\big)
                                     \varphi_{+}(\omega';\sigma',\sigma)
\cr
             +\tr[\proj{R}&\gamma^{\mu}\gamma^{\nu}]
                         \big(\varphi_{-}(\omega;\sigma,\sigma')
                                     -\varphi_{+}(\omega;\sigma,\sigma')\big)
                                    \varphi_{-}(\omega';\sigma',\sigma)\Big\}
\cr}
\cr
&
\eqalign{\sitarel{\rightarrow}{D\rightarrow2}{1\over4\pi}Me^{-M\absig}
   &\{\tr[\proj{L}\gamma^{\mu}\gamma^{\nu}]\delta(\sigma-\sigma')
\cr
    &-\tr[\proj{R}\gamma^{\mu}\gamma^{\nu}]
           (\delta(\sigma-\sigma') - Me^{-M\absig})\}
\cr}
\cr
&\sitarel{\rightarrow}{M\rightarrow\infty}
   {1\over4\pi}\tr[\proj{L}\gamma^{\mu}\gamma^{\nu}]
                                          \delta(\sigma)\delta(\sigma').
&\EF\cr}
$$

   The remaining terms in $\Pi^{\mu\nu}_c$, on the other hand, are
  summed with $\Pi^{\mu\nu}_b$ to yield $\Pi^{\mu\nu}_{CS}$ $+$
  $\Pi^{\mu\nu}_1$ $+$ $\Pi^{\mu\nu}_2$, where
\eqnn\EGCS \eqnn\EGa \eqnn\EGb
$$
\eqalignno{
&\eqalign{
\Pi^{\mu\nu}_{CS}(p;\omega,\omega';\alpha) = -&\DENNO{2-\DT}
\cr
               &\times\pder{\sigma}\Big\{M(\sigma)
                    \big[\tr[\proj{L}\gamma^{\mu}\gamma^{\nu}]
                        \varphi_{+}(\omega;\sigma,\sigma')
                                     \varphi_{+}(\omega';\sigma',\sigma)
\cr
                &\qquad\qquad\qquad-\tr[\proj{R}\gamma^{\mu}\gamma^{\nu}]
                        \varphi_{-}(\omega;\sigma,\sigma')
                                     \varphi_{-}(\omega';\sigma',\sigma)\big]
                                                                  \Big\},
\cr}
&\EGCS\cr
&\eqalign{
\Pi^{\mu\nu}_1(p;\omega,\omega';\alpha)=
          {1 \over (1-{D\over 2})}\alpha\pder{\alpha}&
                 {1 \over [M^2+\alpha\omega^2+(1-\alpha){\omega'}^2
                                 -\alpha(1-\alpha)p^2]^{1-{D\over 2}}}
\cr
            \times\Big\{
              &\tr[\proj{L}\gamma^{\mu}\gamma^{\nu}]
                        \varphi_{+}(\omega;\sigma,\sigma')
                                     \varphi_{+}(\omega';\sigma',\sigma)
\cr
                &+\tr[\proj{R}\gamma^{\mu}\gamma^{\nu}]
                        \varphi_{-}(\omega;\sigma,\sigma')
                                     \varphi_{-}(\omega';\sigma',\sigma)\Big\},
\cr}
&\EGa\cr
&\eqalign{
\Pi^{\mu\nu}_2(p;\omega,\omega';\alpha) = &\DENNO{2-\DT}
\cr
     &\times\bigg[
             \Big\{\tr[\proj{L}\gamma^{\mu}\gamma^{\nu}]
                               \big({1\over2}\dpder{\sigma}+\alpha^2p^2\big)
      -\alpha(1-\alpha)
               \tr[\proj{R}\gamma^{\mu}\lslash{p}\gamma^{\nu}\lslash{p}]\Big\}
\cr
&\hskip 1cm            \times\varphi_{+}(\omega;\sigma,\sigma')
                                     \varphi_{+}(\omega';\sigma',\sigma)
\cr
      &+\Big\{\tr[\proj{R}\gamma^{\mu}\gamma^{\nu}]
                               \big({1\over2}\dpder{\sigma}+\alpha^2p^2\big)
      -\alpha(1-\alpha)
               \tr[\proj{L}\gamma^{\mu}\lslash{p}\gamma^{\nu}\lslash{p}]\Big\}
\cr
&\hskip 1cm            \times\varphi_{-}(\omega;\sigma,\sigma')
                                     \varphi_{-}(\omega';\sigma',\sigma)
                                                                        \bigg].
\cr}
&\EGb
\cr}
$$
  This $\Pi^{\mu\nu}_{CS}$ is the second term in \DIP.

   In order to perform  $\omega$-integration in \EB\ for these
  integrand, we have only to evaluate the following type of integral:
\eqn\EH{
  \widetilde\Phi^{(\kappa)}_{\pm}(p^2; \sigma, \sigma'; \alpha)
   =
     \int{d\omega d\omega' \over (2\pi)^2}
      {\varphi_{\pm}(\omega;\sigma,\sigma')
                            \varphi_{\pm}(\omega';\sigma,\sigma')
                \over \DENO{\kappa-\DT}},
}
  where $\kappa = 1, 2$. This is Taylor-expanded with respect to $p^2$
  as
\eqn\EI{
  \sum^{\infty}_{n=0}{\Gamma(n+\kappa-\DT) \over \Gamma(\kappa-\DT)\ n!}
       (\alpha(1-\alpha)p^2)^n{M\over(M^2)^{n-\DT+\kappa}}
           \Phi_{\pm}^{(n+\kappa)}(\sigma, \sigma'; \alpha),
}
  where
\eqn\EJ{
  \Phi_{\pm}^{(n)}(\sigma, \sigma'; \alpha)
   =
    {(M^2)^{n-\DT}\over M}
     \int{d\omega d\omega' \over (2\pi)^2}
      {\varphi_{\pm}(\omega;\sigma,\sigma')
                            \varphi_{\pm}(\omega';\sigma,\sigma')
                                                        \over \deno{n-\DT}}.
}
  Using the function \EH, we can write
\eqnn\PICS \eqnn\PIa \eqnn\PIb
$$
\eqalignno{
&\eqalign{
  &\int{d\omega d\omega' \over (2\pi)^2}
    \Pi^{\mu\nu}_{CS}(p;\omega,\omega';\alpha)
 \cr
     &=-\pder{\sigma}\Big\{M(\sigma)
         \big[\tr[\proj{L}\gamma^{\mu}\gamma^{\nu}]
          \widetilde\Phi^{(2)}_{+}(p^2; \sigma, \sigma'; \alpha)
 \cr
&\hskip 5cm  -\tr[\proj{R}\gamma^{\mu}\gamma^{\nu}]
          \widetilde\Phi^{(2)}_{-}(p^2; \sigma, \sigma'; \alpha)\big]\Big\},
 \cr}
&\PICS\cr
&\eqalign{
         &\int{d\omega d\omega' \over (2\pi)^2}
        \Pi^{\mu\nu}_{1}(p;\omega,\omega';\alpha)
 \cr
         &={1 \over (1-{D\over 2})}\alpha\pder{\alpha}
            \Big\{
              \tr[\proj{L}\gamma^{\mu}\gamma^{\nu}]
              \widetilde\Phi^{(1)}_{+}(p^2; \sigma, \sigma'; \alpha)
 \cr
&\hskip 5cm  +\tr[\proj{R}\gamma^{\mu}\gamma^{\nu}]
              \widetilde\Phi^{(1)}_{-}(p^2; \sigma, \sigma'; \alpha)\Big\},
 \cr}
&\PIa\cr
&\eqalign{
         &\int{d\omega d\omega' \over (2\pi)^2}
        \Pi^{\mu\nu}_{2}(p;\omega,\omega';\alpha)
 \cr
     &=\bigg[
             \Big\{\tr[\proj{L}\gamma^{\mu}\gamma^{\nu}]
                               \big({1\over2}\dpder{\sigma}+\alpha^2p^2\big)
      -\alpha(1-\alpha)
               \tr[\proj{R}\gamma^{\mu}\lslash{p}\gamma^{\nu}\lslash{p}]\Big\}
 \cr
 &\hskip 1cm       \times\widetilde\Phi^{(2)}_{+}(p^2; \sigma, \sigma'; \alpha)
 \cr
      &+\Big\{\tr[\proj{R}\gamma^{\mu}\gamma^{\nu}]
                               \big({1\over2}\dpder{\sigma}+\alpha^2p^2\big)
      -\alpha(1-\alpha)
               \tr[\proj{L}\gamma^{\mu}\lslash{p}\gamma^{\nu}\lslash{p}]\Big\}
 \cr
 &\hskip 1cm       \times\widetilde\Phi^{(2)}_{-}(p^2; \sigma, \sigma'; \alpha)
                                                                        \bigg].
 \cr}
&\PIb
\cr}
$$

  We introduce another expression for $\varphi_{\pm}(\omega;\sigma,\sigma')$;
\eqn\phfun{
  \varphi_{\pm}(\omega;\sigma,\sigma')
   = \cos\omega(\sigma-\sigma')
        - {M^2 \over \omega^2+M^2}h_{\pm}(\omega; \sigma,\sigma'),
}
  where
\eqn\EK{
\eqalign{
  h_{\pm}(\omega; &\sigma,\sigma')
\cr
   =  &P_{+}(\sigma,\sigma')
         \left[1 \pm \epsilon(\sigma'){1 \over M}\pder{\sigma}\right]
                                                   \cos\omega(\sigma+\sigma')
\cr
     +&P_{-}(\sigma,\sigma')
         \left[1 \mp \epsilon(\sigma'){1 \over M}\pder{\sigma}\right]
                                                   \cos\omega(\sigma-\sigma')
\cr}
}
  with
\eqn\EL{
  P_{\pm}(\sigma,\sigma')
        = {1 \over 2}(1 \pm \epsilon(\sigma)\epsilon(\sigma')).
}
  These $P_{\pm}(\sigma,\sigma')$ have the property of the
  projection operators:
\eqn\EM{
  P_{\pm}(\sigma,\sigma')P_{\pm}(\sigma,\sigma') = P_{\pm}(\sigma,\sigma'),
  \quad {\rm and}\quad
  P_{\pm}(\sigma,\sigma')P_{\mp}(\sigma,\sigma') = 0.
}
  Substituting these expressions into $\Phi_{\pm}^{(n)}$, we have
\eqn\EN{
\eqalign{
  \Phi_{\pm}^{(n)}(\sigma, \sigma'; \alpha)
    =   \Omega_{\pm}^{(n)}(\sigma, \sigma'; \alpha)
       &-C_{\pm}^{(n)}(\sigma, \sigma'; \alpha)
\cr
       &-C_{\pm}^{(n)}(\sigma, \sigma'; 1-\alpha)
        +H_{\pm}^{(n)}(\sigma, \sigma'; \alpha),
\cr}
}
  where
\eqnn\EOa \eqnn\EOb \eqnn\EOc
$$
\eqalignno{
& \Omega_{\pm}^{(n)}(\sigma, \sigma'; \alpha)
    ={(M^2)^{n-\DT}\over M}\int{d\omega d\omega' \over (2\pi)^2}
     {\cos\omega(\sigma-\sigma')\cos\omega'(\sigma-\sigma')
                                                       \over \deno{n-\DT}},
&\EOa\cr
& \eqalign{
  C_{\pm}^{(n)}(\sigma, \sigma'; \alpha)
    ={(M^2)^{n-\DT}\over M}\int{d\omega d\omega' \over (2\pi)^2}
     &{M^2 \over {\omega'}^2+M^2}
\cr
    &\quad\times{h_{\pm}(\omega';\sigma,\sigma')\cos\omega(\sigma-\sigma')
                                                        \over \deno{n-\DT}},
\cr}
&\EOb\cr
& \eqalign{
  H_{\pm}^{(n)}(\sigma, \sigma'; \alpha)
    ={(M^2)^{n-\DT}\over M}\int{d\omega d\omega' \over (2\pi)^2}
     &{M^4 \over ({\omega}^2+M^2)({\omega'}^2+M^2)}
\cr
    &\quad\times{h_{\pm}(\omega;\sigma,\sigma')h_{\pm}(\omega';\sigma,\sigma')
                                                       \over \deno{n-\DT}}.
\cr}
&\EOc
\cr}
$$
  In order to evaluate these functions, we first use the
  Feynman-parameter technique to unify
\eqn\EP{
  {1 \over [{\alpha\omega^2+(1-\alpha){\omega'}^2+M^2}]^{n-{D \over 2}}}
}
  with
\eqn\EQ{
  {1 \over \omega^2+M^2} \qquad{\rm or}\qquad{1 \over {\omega'}^2+M^2}.
}
  Next we exponentiate these factors as usual via the formula
\eqn\AT{
   {1 \over Q^{n}} = {1\over\Gamma(n)}
                         \int^{\infty}_{0} d\tau \tau^{n-1}\exp[-\tau Q],
}
  and perform $\omega$-($\omega'$-)integrations using the formula
\eqn\ER{
\int{d\omega \over 2\pi}e^{-\tau\omega^2}\cos\omega\sigma
          = ({1 \over 4\pi\tau})^{1\over2}e^{-{\sigma^2 \over 4\tau}}.
}
  Then, rescaling $\tau$'s to $(\tau / M^2)$'s, we obtain
  the following results:
\eqnn\ESa \eqnn\ESb \eqnn\ESc
$$
\eqalignno{
 \Omega_{\pm}^{(n)}&(\sigma, \sigma'; \alpha)
    ={1 \over \Gamma(n-\DT)}{(M^2)^{n-\DT}\over M}
            \int^{\infty}_0d\tau \tau^{(n-\DT)-1}e^{-\tau M^2}
\cr
&\hskip 2cm       \times \int{d\omega \over 2\pi}
                     e^{-\tau\alpha\omega^2}\cos\omega(\sigma-\sigma')
                   \int{d\omega' \over 2\pi}
                     e^{-\tau(1-\alpha){\omega'}^2}\cos\omega'(\sigma-\sigma')
\cr
    &={1\over 4\pi}{1 \over \Gamma(n-\DT)}({1 \over\alpha(1-\alpha)})^{1\over2}
       I_{(n-\DT-1)}(\sigma-\sigma';{1 \over\alpha(1-\alpha)}),
&\ESa\cr
  C_{\pm}^{(n)}&(\sigma, \sigma'; \alpha)
    ={(M^2)^{n-{D-1\over2}} \over \Gamma(n-\DT)}
       \int^1_0 d\beta \beta^{n-\DT-1}
       \int^{\infty}_0d\tau \tau^{(n-\DT)}e^{-\tau M^2}
\cr
&\hskip 2cm       \times \int{d\omega' \over 2\pi}
         e^{-\tau(1-\alpha\beta){\omega'}^2}h_{\pm}(\omega'; \sigma,\sigma')
                 \int{d\omega \over 2\pi}
               e^{-\tau\alpha\beta\omega^2}\cos\omega(\sigma-\sigma')
\cr
  &={1 \over 4\pi}{1 \over \Gamma(n-\DT)}\int^1_0 d\beta \beta^{n-\DT-1}
         ({1 \over \alpha\beta(1-\alpha\beta)})^{1\over2}
\cr
  &\quad \times \bigg\{{1\over M}P_{+}(\sigma,\sigma')
          \big[1\pm\epsilon(\sigma')
                       {1 \over 2M}(\pder{\sigma}+\pder{\sigma'})\big]
            J_{(n-\DT)}(\sigma-\sigma',\sigma+\sigma';\alpha\beta)
\cr
  &\qquad\quad+P_{-}(\sigma,\sigma')
          \big[1 \mp \epsilon(\sigma'){\alpha\beta \over M}\pder{\sigma}\big]
            I_{(n-\DT)}(\sigma-\sigma'; {1\over\alpha\beta(1-\alpha\beta)})
                                                                       \bigg\},
&\ESb\cr
  H_{\pm}^{(n)}&(\sigma, \sigma'; \alpha)
\cr
&\eqalign{
  &={(M^2)^{n-{D-3\over2}} \over \Gamma(n-\DT)}
       \int^1_0 d\beta_1d\beta_2\beta^{n-\DT-1}_2
       \int^{\infty}_0d\tau \tau^{(n-\DT)+1}e^{-\tau M^2}
 \cr
  &\qquad \times \int{d\omega \over 2\pi}
               e^{-\tau(1-\beta_1-\beta_2(1-\alpha))\omega^2}
                                     h_{\pm}(\omega; \sigma,\sigma')
                 \int{d\omega' \over 2\pi}
               e^{-\tau(\beta_1+\beta_2(1-\alpha)){\omega'}^2}
                                     h_{\pm}(\omega'; \sigma,\sigma')
 \cr}
\cr
&\eqalign{
  &={1\over4\pi}{1\over \Gamma(n-\DT)}\int^1_0 d\beta_1d\beta_2
     ({1\over f(\alpha,\beta_1,\beta_2)})^{1\over2}\beta^{n-\DT-1}_2
 \cr
  &\quad\times\bigg[P_{+}(\sigma,\sigma')
    \Big\{\big(1\pm\epsilon(\sigma'){1 \over M}\pder{\sigma}
                  +{f(\alpha,\beta_1,\beta_2)\over M^2}\dpder{\sigma}\big)
   I_{(n-\DT+1)}(\sigma+\sigma';{1\over f(\alpha,\beta_1,\beta_2)})
 \cr&\hskip 5cm  +{1\over2}
    I_{(n-\DT)}(\sigma+\sigma';{1\over f(\alpha,\beta_1,\beta_2)})\Big\}
 \cr
  &\qquad+P_{-}(\sigma,\sigma')
    \Big\{\big(1\mp\epsilon(\sigma'){1 \over M}\pder{\sigma}
                  +{f(\alpha,\beta_1,\beta_2)\over M^2}\dpder{\sigma}\big)
   I_{(n-\DT+1)}(\sigma-\sigma';{1\over f(\alpha,\beta_1,\beta_2)})
 \cr&\hskip 5cm  +{1\over2}
    I_{(n-\DT)}(\sigma-\sigma';{1\over f(\alpha,\beta_1,\beta_2)})\Big\}
                                                                       \bigg],
 \cr}
&\ESc
\cr}
$$
  with $f(\alpha,\beta_1,\beta_2)=$ $(\beta_1+\beta_2(1-\alpha))$
  $(1-\beta_1-\beta_2(1-\alpha))$.

   From these equations \ESa, \ESb, and \ESc, we see that,
  for $\Phi_{\pm}^{(n)}(\sigma, \sigma'; \alpha)$ with $n\geq2$,
  we can remove the regularization \IE\ $D\rightarrow2$ and then
  let $M$ go to infinity. Such a limit is seen from Appendix A to be
\eqn\phLIM{
  \Phi_{\pm}^{(n)}(\sigma, \sigma'; \alpha)|_{D=2}
   \sitarel{\rightarrow}{M\rightarrow\infty}
   {1 \over (4\pi)^{{1\over 2}}}{\Gamma(n-{3\over 2})\over\Gamma(n-1)}
                                                  \delta(\sigma-\sigma').
     \quad(n\geq2)
}
  Therefore, the $O(p^2)$ terms  in \PICS, \PIa, and \PIb\ cannot
  contribute in the large mass limit, and the terms proportional to
  $\dpder{\sigma}\widetilde\Phi^{(2)}_{\pm}(p^2; \sigma, \sigma'; \alpha)$
  cannot in \PIb, either. So $\Pi^{\mu\nu}_2$ vanishes in the
  large mass limit.

   The remaining part in \PICS\ can contribute to
  $\Pi^{\mu\nu}_D(p;\sigma,\sigma')$ in the large mass limit as
\eqn\ET{
\eqalign{
     &-{M\over4\pi}\int^1_0d\alpha\pder{\sigma}\Big\{\epsilon(\sigma)
         \big[\tr[\proj{L}\gamma^{\mu}\gamma^{\nu}]
          \Phi^{(2)}_{+}(\sigma, \sigma'; \alpha)
             -\tr[\proj{R}\gamma^{\mu}\gamma^{\nu}]
          \Phi^{(2)}_{-}(\sigma, \sigma'; \alpha)\big]\Big\}
\cr
     &\sitarel{\rightarrow}{M\rightarrow\infty}
      {1\over4\pi}\pder{\sigma}\Big\{\epsilon^{\mu\nu}\epsilon(\sigma)
       \delta(\sigma-\sigma')\Big\}.
\cr}
}
  This is the second term in \diracpi.

   Finally, we examine the term with $\Phi^{(1)}_{\pm}$:
\eqn\EU{
         {M \over (1-{D\over 2})}\alpha\pder{\alpha}
            \Big\{
              \tr[\proj{L}\gamma^{\mu}\gamma^{\nu}]
              \Phi^{(1)}_{+}(\sigma, \sigma'; \alpha)
              +\tr[\proj{R}\gamma^{\mu}\gamma^{\nu}]
              \Phi^{(1)}_{-}(\sigma, \sigma'; \alpha)\Big\}
}
  in \PIa. Note that we can set $D$ equal to $2$ in \EU\ thanks to
  the factor ${1 \over \Gamma(1-\DT)}$ contained in $\Phi^{(1)}_{\pm}$.
  If we perform the $\alpha$-differentiation in \EU\ using the
  relations \AOa\ and \AOb\ in Appendix~B, we find that all the terms
  in $\Phi^{(1)}_{\pm}$, except one type of terms, yield the factor
  ${1\over M^2}$ and give no contributions in the large mass limit.
  The exceptional terms are
\eqn\EVa{
  \mp{\alpha\over4\pi}P_{-}(\sigma,\sigma')\epsilon(\sigma')\pder{\sigma}
  \int^1_0d\beta \Big({1\over \alpha\beta(1-\alpha\beta)}\Big)^{1\over2}
            I_0(\sigma-\sigma';{1 \over \alpha\beta(1-\alpha\beta)})
}
  in ${M \over (1-\DT)}\alpha\pder{\alpha}C^{(1)}_{\pm}(\sigma,\sigma';\alpha)$
  and
\eqn\EVb{
 \eqalign{
  \pm{\alpha\over4\pi}P_{-}(\sigma,\sigma')\epsilon(\sigma')\pder{\sigma}
 \int^1_0d\beta &
             \Big({1\over (1-\alpha)\beta(1-(1-\alpha)\beta)}\Big)^{1\over2}
 \cr
  &\times I_0(\sigma-\sigma';{1 \over (1-\alpha)\beta(1-(1-\alpha)\beta)})
 \cr}
}
  in ${M\over(1-\DT)}\alpha\pder{\alpha}C^{(1)}_{\pm}
  (\sigma,\sigma';1-\alpha)$. In the large mass limit, however,
  these terms cancel each other in \EU\ and do not contribute, either.
  The term \EU\  thus vanishes in the large mass limit.

   In summary, only the terms from \EF\ and \ET\ can contribute to the
  pure Dirac-type contribution $\Pi^{\mu\nu}_D(p;\sigma,\sigma')$
  in the large mass limit. Thus,
\eqn\EW{
\Pi^{\mu\nu}_D(p;\sigma,\sigma')
   \sitarel{\rightarrow}{M \rightarrow \infty}
         ({1 \over 4\pi})\left[
                    (g^{\mu\nu}-\epsilon^{\mu\nu})
                                             \delta(\sigma)\delta(\sigma')
                    +\epsilon^{\mu\nu}\pder{\sigma}
                              \big(\epsilon(\sigma)\delta(\sigma-\sigma')\big)
                                                                      \right].
}
  This is \diracpi.

\listrefs

\end